\documentclass[
    aps,
    prd,
    reprint,
    nolongbibliography,
    nobibnotes,
    nofootinbib,
]{revtex4-2}

\usepackage[dvipsnames, usenames]{xcolor}
\definecolor{linkcolor}{rgb}{0.1, 0.5, 0.7}

\usepackage[
    colorlinks=true,
    linkcolor=linkcolor,
    citecolor=linkcolor,
    filecolor=linkcolor,
    urlcolor=linkcolor,
    pdfusetitle,
]{hyperref}

\usepackage{orcidlink}
\usepackage{amssymb}
\usepackage{amsmath}
\usepackage{physics}

\newcommand{\comment}[1]{}

\newcommand{\LIGO}{\affiliation{LIGO Laboratory, Massachusetts Institute of Technology, Cambridge, MA 02139, USA}}
\newcommand{\MKI}{\affiliation{Kavli Institute for Astrophysics and Space Research, Massachusetts Institute of Technology, Cambridge, MA 02139, USA}}
\newcommand{\MIT}{\affiliation{Department of Physics, Massachusetts Institute of Technology, Cambridge, MA 02139, USA}}

\begin{document}

\title{Rapid inference and comparison of gravitational-wave population models with neural variational posteriors}

\author{Matthew Mould\,\orcidlink{0000-0001-5460-2910}}
\email{mmould@mit.edu}
\LIGO\MKI\MIT

\author{Noah E. Wolfe\,\orcidlink{0000-0003-2540-3845}}
\LIGO\MKI\MIT

\author{Salvatore Vitale\,\orcidlink{0000-0003-2700-0767}}
\LIGO\MKI\MIT

\date{\today}

\begin{abstract}
The LIGO--Virgo--KAGRA catalog has been analyzed with an abundance of different population models due to theoretical uncertainty in the formation of gravitational-wave sources. To expedite model exploration, we introduce an efficient and accurate variational Bayesian approach that learns the population posterior with a normalizing flow and serves as a drop-in replacement for existing samplers. With hardware acceleration, inference takes just seconds for the current set of black-hole mergers and readily scales to larger catalogs. The trained posteriors provide an arbitrary number of independent samples with exact probability densities, unlike established stochastic sampling algorithms, while requiring up to three orders of magnitude fewer likelihood evaluations and as few as $\mathcal{O}(10^3)$. Provided the posterior support is covered, discrepancies can be addressed with smoothed importance sampling, which quantifies a goodness-of-fit metric for the variational approximation while also estimating the evidence for Bayesian model selection. Neural variational inference thus enables interactive development, analysis, and comparison of population models, making it a useful tool for astrophysical interpretation of current and future gravitational-wave observations.
\end{abstract}

\maketitle

\section{Introduction}
\label{sec: Introduction}

The catalog of gravitational-wave (GW) sources \cite{LIGOScientific:2018mvr, LIGOScientific:2020ibl, LIGOScientific:2021usb, KAGRA:2021vkt} observed by the LIGO--Virgo--KAGRA (LVK) detectors \cite{LIGOScientific:2014pky, VIRGO:2014yos, KAGRA:2020tym} has become a test bed for models of compact-binary formation. Binary black-hole (BH) mergers have received particular focus for population studies \cite{LIGOScientific:2018jsj, LIGOScientific:2020kqk, KAGRA:2021duu} as they comprise the largest sample of detections. Such analyses require a model for the rate (equivalently, the number or probability density) of mergers across the source parameter space, e.g., masses, spins, and redshift \cite{Mandel:2018mve, Vitale2020}. However, due to significant theoretical uncertainties surrounding the formation of merging BHs \cite{Mapelli2020, Mandel:2021smh, Mandel:2018hfr, Belczynski:2021zaz}, it has become most common to use and compare different phenomenological models (see, e.g., Ref.~\cite{Callister:2024cdx} and references therein). Designing such models requires analyzing the same data many times with slight variations to discern the best fit, introducing a significant analysis burden that only increases as the catalog grows, which recent work has aimed to curtail \cite{Fabbri:2025faf}.

We propose variational inference \cite{Blei03042017} as a rapid and accurate method for GW population inference. The basic idea is to convert the intractable Bayesian posterior from a sampling problem (as in Markov chain Monte Carlo or nested sampling \cite{Metropolis:1953am, Hastings:1970aa, Skilling:2004pqw, Skilling:2006gxv}) to an optimization problem by assuming a parametrized distribution that approximates the true posterior. With a sufficiently flexible approximation such as a normalizing flow \cite{Papamakarios:2019fms, Kobyzev:2019ydm}, we show that it is possible to match the results of state-of-the-art inference algorithms at a fraction of the computation time. Independent of the catalog size, our method requires as few as $\mathcal{O}(10^3$--$10^4)$ likelihood evaluations, resulting in predictable computation time based on the cost of a single likelihood evaluation. Moreover, the learned posteriors are exact probability density functions that can generate an arbitrary number of samples guaranteed to be statistically independent, which can in turn can be used to estimate the Bayesian evidence via neural importance sampling \cite{10.1145/3341156, Dax:2022pxd}. This allows us to assess the predictions of population models fitted to GW data and select between them. A smoothed variant of importance sampling \cite{JMLR:v25:19-556} can be used to regularize reweighted samples generated from the learned posterior if necessary and produces a goodness-of-fit metric that indicates when the variational approximation should be improved (e.g., through longer training or a more flexible posterior model), without needing a comparison to results from independent sampling algorithms.

On current GW catalogs, accurate neural variational posteriors can be trained in a matter of seconds using GPU hardware acceleration. This allows population models to be tested interactively on real data, thereby greatly expediting the process of model exploration and astrophysical interpretation. We further show that a mock catalog containing 1599 events simulated with full Bayesian parameter estimation can be analyzed accurately in minutes (instead of several hours for nested sampling), implying variational inference will be a useful tool for upcoming LVK observing runs.

We describe the details of variational inference in Sec.~\ref{sec: Variational inference}, demonstrate its efficacy for GW population analyses on current and future catalogs in Sec.~\ref{sec: Gravitational-wave populations}, and conclude in Sec.~\ref{sec: Discussion}. Code to reproduce our results is available in Ref.~\cite{matthew_mould_2024_12770127}\footnote{\href{https://github.com/mdmould/gwax}{github.com/mdmould/gwax}}.

\section{Variational inference}
\label{sec: Variational inference}

\subsection{Posterior approximation}
\label{sec: Posterior approximation}

The goal of Bayesian inference is to measure parameters $\lambda$ that are not directly observable from observed data. According to Bayes' theorem, this measurement is given by the posterior distribution $\mathcal{P}(\lambda) = \mathcal{L}(\lambda) \pi(\lambda) / \mathcal{Z}$, where $\mathcal{L}(\lambda)$ is the likelihood function, $\pi(\lambda)$ is the prior, and $\mathcal{Z} = \int \dd{\lambda} \mathcal{L}(\lambda) \pi(\lambda)$ is the evidence. Typically, the joint density $\mathcal{L}(\lambda) \pi(\lambda)$ can be evaluated by assumption, but the evidence integral---and therefore the posterior---is intractable. Instead, numerical algorithms such as Markov chain Monte Carlo and nested sampling are commonly used in GW astronomy \cite{Veitch:2014wba, Ashton:2018jfp, Romero-Shaw:2020owr} to draw samples from the posterior (and potentially estimate the evidence); those samples then feed into downstream predictive modeling, e.g., to estimate the underlying population properties of BH mergers, e.g., Refs.~\cite{KAGRA:2021duu, Mandel:2018mve, Vitale2020, Callister:2024cdx}.

Variational inference is a Bayesian approach that assumes the posterior can be approximated by a parametric distribution $\mathcal{Q}(\lambda|\varphi)$ \cite{Blei03042017}. We seek the variational parameters $\varphi$ such that the approximation $\mathcal{Q}$ most closely matches the truth $\mathcal{P}$. This requires a functional notion of similarity between probability distributions; there is no unique choice, but for computational tractability a common one is the Kullback--Leibler (KL) divergence \cite{Kullback:1951zyt}
\begin{align}
\mathrm{KL} [ \mathcal{Q} , \mathcal{P} ]
:=
\int \dd{\lambda}
\mathcal{Q} ( \lambda | \varphi )
\ln \frac
{ \mathcal{Q} ( \lambda | \varphi ) }
{ \mathcal{P} ( \lambda ) }
\, .
\label{eq: kl}
\end{align}
As $\lambda$ may be high dimensional, Eq.~(\ref{eq: kl}) is most easily approximated with Monte Carlo integration. Hence, $\mathrm{KL}[\mathcal{Q},\mathcal{P}]$ is used over $\mathrm{KL}[\mathcal{P},\mathcal{Q}]$ because we can draw samples from $\mathcal{Q}$ and evaluate their probability density by choosing an appropriate variational family, but we cannot do the same for $\mathcal{P}$. Using Bayes theorem, we define an equivalent loss function
\begin{align}
L(\varphi) :=
\mathrm{KL} [ \mathcal{Q} , \mathcal{P} ] - \ln\mathcal{Z}
\approx
\frac{1}{B} \sum_{i=1}^B
\ln \frac
{ \mathcal{Q}(\lambda_i|\varphi) }
{ \mathcal{L}(\lambda_i) \pi(\lambda_i) }
\, ,
\label{eq: loss}
\end{align}
where likelihood evaluations $\mathcal{L}(\lambda_i)$ are vectorized across a batch of $B$ samples $\{ \lambda_i \}_{i=1}^B \sim \mathcal{Q}(\cdot|\varphi)$. Assuming the target joint distribution (and in particular the likelihood function) is differentiable, the loss function is optimized to find $\hat{\varphi} = \mathrm{argmin}_\varphi L(\varphi)$ by computing $\nabla_\varphi L(\varphi)$ via automatic differentiation and performing stochastic gradient descent \cite{Kingma:2013hel, JMLR:v18:16-107}.

\subsection{Importance sampling and evidence estimation}
\label{sec: Importance sampling and evidence estimation}

Note that $\mathrm{KL}[\mathcal{Q},\mathcal{P}] \geq 0$ implies a lower bound on the evidence, $\ln\mathcal{Z} \geq -L(\varphi)$ \cite{Kingma:2013hel}. If $\hat{\mathcal{Q}}(\lambda) := \mathcal{Q}(\lambda|\hat{\varphi}) \approx \mathcal{P}(\lambda)$ after training, the evidence can be estimated by importance sampling $M$ draws $\{\lambda_i\}_{i=1}^M$ from $\hat{\mathcal{Q}}$ \cite{10.1145/3341156, Dax:2022pxd}:
\begin{align}
\mathcal{Z} =
\int \dd{\lambda} \hat{\mathcal{Q}}(\lambda)
\frac { \mathcal{L}(\lambda) \pi(\lambda) }
{ \hat{\mathcal{Q}}(\lambda) }
\approx
\frac{1}{M} \sum_{i=1}^M
\frac { \mathcal{L}(\lambda_i) \pi(\lambda_i) }
{ \hat{\mathcal{Q}}(\lambda_i) }
\, .
\label{eq: evidence}
\end{align}
When training variational posteriors for multiple models, the evidences can be used to compute Bayes factors for model comparison. The importance ratios $w_i := \mathcal{L}(\lambda_i) \pi(\lambda_i) / \hat{\mathcal{Q}}(\lambda_i)$ can be used to correct discrepancies between $\hat{\mathcal{Q}}$ and $\mathcal{P}$ by applying weights $\propto w_i$ to the samples $\lambda_i \sim \hat{\mathcal{Q}}$. The importance sampling efficiency $\varepsilon = \left( \sum_i w_i \right)^2 / \left( M \sum_i w_i^2 \right) \in [0,1]$ will be close to unity (zero) if $\hat{\mathcal{Q}}$ is a good (bad) approximation, with an effective sample size $\varepsilon M$ \cite{kong1992note, elvira2022rethinking}. The variance of this estimator is given by $\mathbb{V}[\ln\mathcal{Z}] \approx (1/\varepsilon - 1) / M$. By assumption, importance sampling requires $\mathrm{supp}\,\hat{\mathcal{Q}} \supseteq \mathrm{supp}\,\mathcal{P}$ so that the importance ratio does not diverge. If this condition is met, Eq.~(\ref{eq: evidence}) provides an unbiased estimate for the evidence $\mathcal{Z}$ (this is not true for $\ln\mathcal{Z}$, but we found corrections negligible in practice; see, e.g., Eq.~(7) in the Supplemental Material of Ref.~\cite{Dax:2022pxd}). However, unlike the forward divergence $\mathrm{KL}[\mathcal{P},\mathcal{Q}]$, minimizing the reverse divergence $\mathrm{KL}[\mathcal{Q},\mathcal{P}]$ does not provide this guarantee, which can lead to biased evidence estimation even if $\varepsilon \approx 1$ (e.g., when the variational distribution fits only one of two well-separated posterior modes). Light tails in $\hat{\mathcal{Q}}$ with respect to $\mathcal{P}$ can lead to large importance weights and numerically unstable Monte Carlo integration.

We propose Pareto smoothed importance sampling \cite{JMLR:v25:19-556} to assess the variational fit \cite{pmlr-v80-yao18a}, using (a slightly modified version of\footnote{\href{https://github.com/mdmould/arviz}{github.com/mdmould/arviz}}) the implementation from \textsc{ArviZ} \cite{arviz_2019}. In short, a generalized Pareto distribution is fit \cite{cdfd776c-0cd3-3d1c-b2c4-2f29a751e40b} to a subset (of predetermined size) of the largest importance weights, which are then replaced with their expected values according to the fitted distribution before substitution into Eq.~(\ref{eq: evidence}); see Algorithm~1 of Ref.~\cite{JMLR:v25:19-556}. Reasonable convergence is indicated by a fitted Pareto shape parameter $\hat{k}<0.7$ \cite{JMLR:v25:19-556}; otherwise, the variational posterior should be improved either with more training or a more flexible distribution. In our experiments, we typically found the trained posteriors $\hat{\mathcal{Q}}$ themselves accurately matched results from nested sampling, but compared to results reweighted by the raw importance ratios and their smoothed versions, using $\hat{k}$ as a useful convergence diagnostic.

\subsection{Normalizing flows}
\label{sec: Normalizing flows}

The accuracy of the posterior approximation depends on the choice of the variational distribution. Normalizing flows are probability distributions that map a simple known distribution to a target distribution with a differentiable and invertible transformation using the change-of-variables formula \cite{Papamakarios:2019fms, Kobyzev:2019ydm}. They are parametrized by neural networks which enables them to fit complicated high-dimensional posteriors \cite{pmlr-v37-rezende15}. In this work, we use block neural autoregressive flows \cite{pmlr-v115-de-cao20a, pmlr-v80-huang18d} to transform isotropic standard normal base distributions to our target posteriors via bijective neural networks. We use the implementation and default settings in \textsc{FlowJAX} \cite{ward2023flowjax} and train with Adam optimization \cite{Kingma:2014vow}. The \textsc{JAX} framework allows us to leverage automatic differentiation, GPU hardware acceleration, and just-in-time (JIT) compilation \cite{jax2018github, deepmind2020jax, kidger2021equinox}. To match the support of $\mathcal{Q}$ to that of the bounded prior $\pi$ (and thus $\mathcal{P}$), we additionally constrain the flow output with the standard logistic function, followed by fixed affine transformations to rescale each dimension of $\lambda$.

\begin{figure}
\centering
\includegraphics[width=0.9\columnwidth]{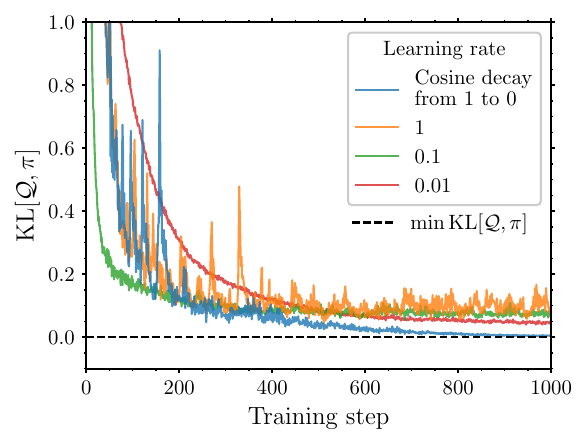}
\caption{Loss functions from training a normalizing flow $\mathcal{Q}$ with variational inference to learn a 14-dimensional uniform prior $\pi$ with different learning rates (colored lines). A cosine schedule (blue) decaying the learning rate from an initial large value of 1 to 0 over $10^3$ training steps induces superconvergence to the minimum KL divergence, $\mathrm{KL}[\mathcal{Q},\pi] = 0$ (dashed black line).}
\label{fig: losses}
\end{figure}

Rather than randomly initializing the neural network parameters $\varphi$, we first match $\mathcal{Q}$ to the prior $\pi$ by minimizing $\mathrm{KL}[\mathcal{Q},\pi]$, which resulted in better efficiencies $\varepsilon$. We show example loss curves of Eq.~(\ref{eq: loss}) from this prior training in Fig.~\ref{fig: losses}. The prior $\pi$ is an independent uniform distribution on a 14-dimensional parameter vector $\lambda$ for the population model considered in Sec.~\ref{sec: Likelihood and model}. We use a variational batch size $B=10^4$ and train for $10^3$ steps, which takes $<10\,\mathrm{s}$ including JIT compilation. We test constant learning rates of 1, 0.1, and 0.01. Instead, using a cosine schedule to decay the learning rate from 1 to 0, we observed superconvergence \cite{10.1117/12.2520589}, where the model reaches a near optimal solution $\mathrm{KL}[\hat{\mathcal{Q}},\pi] \approx 0$ over much fewer training steps. The initial large learning rate allows the model to make large updates and jump between several local minima, but this also prevents local convergence; after global exploration, the decaying learning rate aids the model in converging within the best found local minimum \cite{zhang2019cyclical}. We therefore use this setup also when training the posteriors and observed a similar phenomenon, but using a smaller batch size due to memory limitations, and with an initial learning rate of 0.1 and gradient clipping for stability.

\section{Gravitational-wave populations}
\label{sec: Gravitational-wave populations}

We apply neural variational inference for GW population analysis, focusing on binary BHs. First we describe the likelihood function and regularization, our population model, and a benchmark inference method to compare against our variational posteriors. Then we analyze the current catalog of real GW events and a mock catalog containing a large number of simulated events.

\subsection{Likelihood and model}
\label{sec: Likelihood and model}

When marginalized over a scale-invariant prior \cite{Fishbach:2018edt, Essick:2023upv}, the likelihood for a catalog of $N$ observed GW events that we input to Eq.~(\ref{eq: loss}) is given by
\begin{align}
\mathcal{L}(\lambda) \propto
\prod_{n=1}^N
\frac
{ \int \dd{\theta_n} \mathcal{L}_n(\theta_n) p(\theta_n|\lambda) }
{ \int \dd{\theta} P(\mathrm{det}|\theta) p(\theta|\lambda) }
\, ,
\label{eq: likelihood}
\end{align}
where $\theta$ represents the GW source parameters, $\mathcal{L}_n$ is the GW likelihood function for an individual event labeled $n$, $P(\mathrm{det}|\theta)$ is the probability of detecting a source with parameters $\theta$, and $p(\theta|\lambda)$ is a model for the underlying population parametrized by $\lambda$ --- the parameters we want to infer \cite{Mandel:2018mve, Vitale2020}. We model the distributions of primary BH mass $m_1$ (the heavier BH in the binary) and binary mass ratio $q\in(0,1]$ with the \textsc{Power Law + Peak} model \cite{Talbot:2018cva}, but unlike in Refs.~\cite{Talbot:2018cva, KAGRA:2021duu} we add a smooth turnoff at high masses using the same functional form as the low-mass turn on. Dimensionless BH spin magnitudes $\chi_i\in[0,1)$ ($i=1,2$) are independent and identically distributed (IID) from nonsingular beta distributions \cite{Wysocki:2018mpo}, while spin tilts $\tau_i\in[0,\pi]$ follow a mixture between IID truncated normal distributions in $\cos\tau_i$ peaking at spin alignment ($\cos\tau_i=1$) and an isotropic component \cite{Vitale:2015tea, Talbot:2017yur}. The merger-rate density in the comoving frame evolves over redshift $z$ as $R(z) = R_0 (1+z)^\gamma$, with local ($z=0$) rate $R_0$ \cite{Fishbach:2018edt}. With the log-uniform prior for $R_0$ used to derive the marginal likelihood $\mathcal{L}(\lambda)$ in Eq.~(\ref{eq: likelihood}), the joint posterior $\mathcal{P}(R_0,\lambda) = \mathcal{P}(R_0|\lambda) \mathcal{P}(\lambda)$ can be recovered with a closed-form expression for $\mathcal{P}(R_0|\lambda)$ as a postprocessing step \cite{Fishbach:2018edt}, so we fit for just $\mathcal{P}(\lambda)$. Altogether, $\lambda$ has 14 free parameters for which we take uniform priors.

The integrals in Eq.~(\ref{eq: likelihood}) are typically computed with Monte Carlo approximations (though see Ref.~\cite{Mancarella:2025uat} for an alternative approach), but this introduces additional statistical uncertainty where the likelihood is poorly approximated. A common approach to avoid parameter regions of potentially high bias is setting the likelihood to zero when the effective sample sizes or associated variances of the Monte Carlo integrals pass some chosen threshold \cite{Farr:2019rap, Essick:2022ojx, Talbot:2023pex, KAGRA:2021duu} (though one should bear in mind that this assumes an importance-sampling convergence criterion that itself depends on the accuracy of an estimate obtained by importance sampling \cite{chatterjee2018sample, JMLR:v25:19-556}). We follow Ref.~\cite{Talbot:2023pex} and threshold the likelihood on the estimated Monte Carlo variance $\mathcal{V}(\lambda)$ of $\ln\mathcal{L}(\lambda)$. During variational inference training, however, we instead taper the likelihood with a steep but smooth function given by $\ln\mathcal{T}(\lambda) = -100 \left( \mathcal{V}(\lambda) - V \right)^2$ and replace $\mathcal{L}(\lambda) \mapsto \mathcal{T}(\lambda) \mathcal{L}(\lambda)$ when $\mathcal{V}(\lambda) > V$ \cite{Heinzel:2024jlc}, thereby approximating a threshold at $\mathcal{V}(\lambda) = V$ while preserving the ability for automatic differentiation. After training, the strict threshold is reintroduced in the importance sampling step, such that the overall sampling efficiency $\varepsilon$ is the combination of the variational approximation and weights zeroed by the likelihood threshold. We consider this threshold as a modification $\mathcal{L}(\lambda) \mapsto \mathcal{L}'(\lambda) := \Theta ( V - \mathcal{V}(\lambda) ) \mathcal{L}(\lambda) \pi(\lambda)$ to the likelihood function itself and therefore we use
$\mathcal{Z}' := \int \dd{\lambda} \mathcal{L}'(\lambda) \pi(\lambda)$ when quoting evidences, where $\Theta$ is the Heaviside step function.

As a benchmark for variational inference, we compare to nested sampling using \textsc{Dynesty} \cite{Speagle:2019ivv} through \textsc{Bilby} \cite{Ashton:2018jfp}. For a fair comparison, for both nested sampling and variational inference we use the implementation of the likelihood and population models in \textsc{GWPopulation} \cite{Talbot:2024yqw} and run on a NVIDIA A30 GPU using \textsc{JAX} \cite{jax2018github}. We compare the number of likelihood evaluations required, run times (including JIT compilation and prior training), and evidence estimates with associated uncertainties from nested sampling and the intrinsic Monte Carlo variance from importance sampling (see Sec.~\ref{sec: Importance sampling and evidence estimation}). We also define an efficiency as the effective number of posterior samples per likelihood evaluation, where for nested sampling the effective sample size is computed using the weighted nested samples and for variational inference the importance sampling efficiency is used (see Sec.~\ref{sec: Importance sampling and evidence estimation}).

We use the \textsc{Bilby} default sampler settings for \textsc{Dynesty}, but in the course of this work were made aware of more efficient alternate settings, which we compare to for a more competitive timing benchmark  (in particular, we use \texttt{sample=acceptance-walk} and \texttt{naccept=10}). Note that due to the likelihood threshold $\mathcal{V}(\lambda) > V$, the integral returned by the nested sampling algorithm is actually $\mathcal{Z}' / \int \dd{\lambda} \Theta \big( V - \mathcal{V}(\lambda) \big) \pi(\lambda)$; we estimate and remove this denominator with a simple Monte Carlo approximation, whose uncertainty we estimate and propagate alongside that from nested sampling (though the total uncertainty in $\mathcal{Z}'$ is dominated by the latter).

\subsection{Current catalog}
\label{sec: Current catalog}

In our first analysis, we perform population inference on the LVK catalog of BH mergers from GWTC-2.1 and 3 \cite{LIGOScientific:2021usb, KAGRA:2021vkt}, including the same 69 events with false-alarm rates (FARs) $<1\,\mathrm{yr}^{-1}$ as in Ref.~\cite{KAGRA:2021duu}. We use the \texttt{Mixed} parameter-estimation samples ($\approx2\times10^3$ per event) and estimate the search sensitivity using the ($\approx4\times10^4$) detected binary BH injections from the public data \cite{KAGRA:2023pio}. Prior ranges for the population parameters $\lambda$ are enumerated in Table.~\ref{tab: current priors} (mostly following Appendix B of Ref.~\cite{KAGRA:2021duu}). We limit the variance of $\ln\mathcal{L}(\lambda)$ at a threshold $V=1$ \cite{Talbot:2023pex}. For variational inference, we train for $10^4$ steps with a batch size $B=1$ [Eq.~(\ref{eq: loss})], which we surprisingly found to be sufficient for well-converged posteriors.

\begin{table}
\centering
\setlength{\tabcolsep}{8pt}
\renewcommand*{\arraystretch}{1.1}
\begin{tabular}{ccc}
\hline 
\textbf{Parameter} & \textbf{Description} & \textbf{Range} \\
\hline
$\alpha_m$ & $m_1$ spectral index & $[-12,4]$ \\
$\beta_q$ & $q$ spectral index & $[-2,7]$ \\
$f_m$ & $m_1$ Gaussian fraction & $[0,1]$ \\
$\mu_m$ & $m_1$ Gaussian location [$M_\odot$] & $[20,50]$ \\
$\sigma_m$ & $m_1$ Gaussian width [$M_\odot$] & $[1,10]$ \\
$m_\mathrm{min}$ & Minimum BH mass [$M_\odot$] & $[2,6]$ \\
$m_\mathrm{max}$ & Maximum BH mass [$M_\odot$] & $[70,100]$ \\
$\delta_\mathrm{min}$ & Low-mass smoothing [$M_\odot$] & $[0,10]$ \\
$\delta_\mathrm{max}$ & High-mass smoothing [$M_\odot$] & $[0,10]$ \\
$\alpha_\chi$ & $\chi_i$ beta distribution shape $\alpha$ & $[1,10]$ \\
$\beta_\chi$ & $\chi_i$ beta distribution shape $\beta$ & $[1,10]$ \\
$f_\tau$ & Aligned-spin fraction & $[0,1]$ \\
$\sigma_\tau$ & Aligned-spin $\cos\tau_i$ width & $[0.1,4]$ \\
$\gamma$ & $z$ spectral index & $[-6,6]$ \\
\hline
\end{tabular}
\caption{Parameters of the population model and the range of the uniform priors we use for inference on the current LVK catalog of binary BH mergers with $\mathrm{FAR} < 1 \, \mathrm{yr}^{-1}$.}
\label{tab: current priors}
\end{table}

\begin{figure}
\centering
\includegraphics[width=0.9\columnwidth]{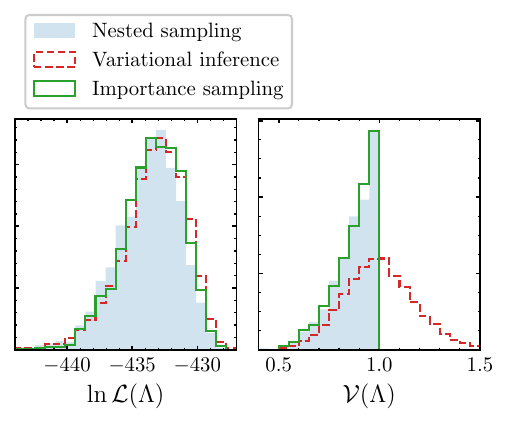}
\caption{The posterior distribution of log-likelihood values $\ln\mathcal{L}(\lambda)$ (left) and associated variances $\mathcal{V}(\lambda)$ (right) from a population analysis of the current LVK catalog of BH mergers inferred using nested sampling (shaded blue), variational inference (dashed red), and variational inference with smoothed importance sampling (solid green).}
\label{fig: extras}
\end{figure}

\begin{figure*}
\includegraphics[width=\textwidth]{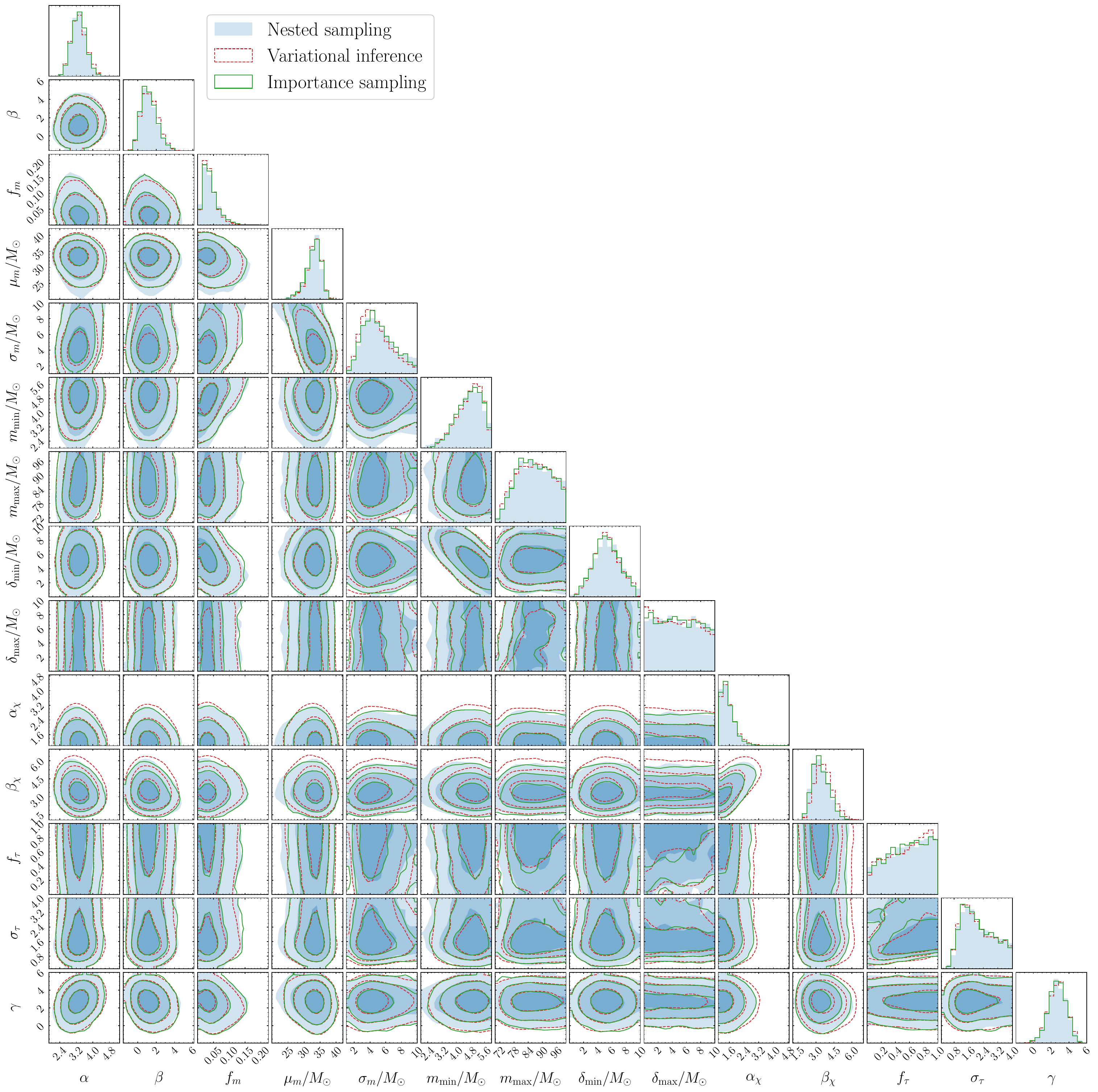}%
\llap{\raisebox{7.6cm}{%
    \includegraphics[width=0.6\textwidth]{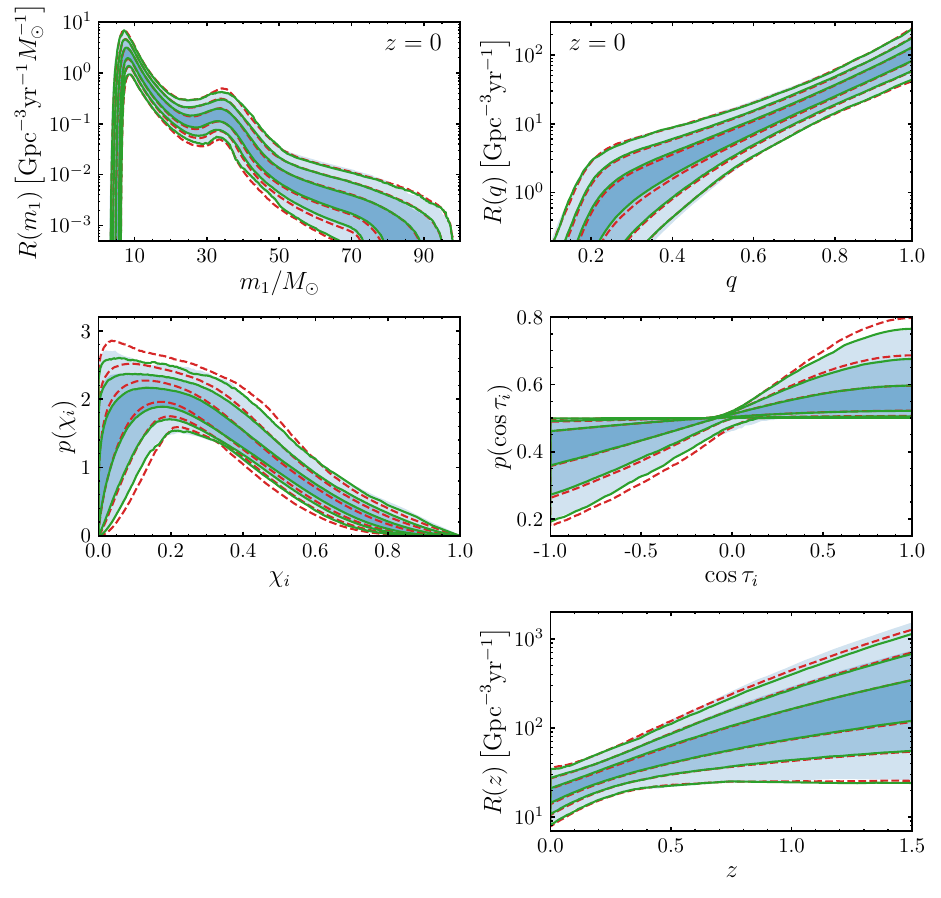}%
}}
\caption{Posterior distributions of population (lower left) and BH source (upper right) parameters, inferred using nested sampling (shaded blue) and variational inference with (solid green) or without (dashed red) smoothed importance sampling. Off-diagonal and upper-right panels show 50\%, 90\%, and 99\% credible regions. For primary mass $m_1$ and mass ratio $q$ (redshift $z$), we show local (evolving) merger rate densities. For BH spin magnitudes $\chi_i$ and tilts $\tau_i$, we show probability densities.}
\label{fig: current}
\end{figure*}

In Fig.~\ref{fig: extras}, we show the posterior distributions of likelihoods given by $\ln\mathcal{L}(\lambda)$ and associated variances $\mathcal{V}(\lambda)$ as inferred using nested sampling, as well as variational inference with and without smoothed importance sampling. Posterior draws from both nested sampling runs were equivalent within finite sampling uncertainty, so we plot results using only the default settings in Fig.~\ref{fig: extras} (and Fig.~\ref{fig: current}). The distributions are in good agreement. Though the variance-based tapering function $\mathcal{T}(\lambda)$ used for variational inference is not completely effective, as $\approx50\%$ of samples have $\mathcal{V}(\lambda)>V$, this is corrected in postprocessing to enforce the threshold of $V=1$. This results in an overall importance sampling efficiency $\varepsilon\approx25$--30\% (with or without Pareto smoothing). The fitted Pareto shape parameter is $\hat{k}\approx0.32$, well within the regime of convergence $<0.7$ recommended in Ref.~\cite{JMLR:v25:19-556}.

In Table~\ref{tab: current stats} we present more detailed results. With the default settings, nested sampling required $>10^6$ likelihood evaluations to reach convergence, or $>3\times10^5$ with the alternate settings. These runs took 10--30\,min, which is significantly faster than running on a CPU without \textsc{JAX} (for comparison, in previous work \cite{Mould:2022xeu} we found run times at the upper end of this range when multiprocessing over $\approx100$ CPU cores). On the other hand, we found that for GWTC-3 we could train accurate variational posteriors with $\mathcal{O}(10^4)$ likelihood evaluations. That this is possible with a batch size of just $B=1$ means that variational inference is possible even when the likelihood carries a large memory cost that prevents larger vectorized batches. Training took $\lesssim50\,\mathrm{s}$ in total, broken up into $\lesssim10\,\mathrm{s}$ for the prior training step (and its JIT compilation), $\approx25\,\mathrm{s}$ for JIT compilation, and just $\approx15\,\mathrm{s}$ for actual posterior training.

Once trained, drawing samples from the variational posterior has negligible cost. The (smoothed) importance reweighting step took an additional $\approx5\,\mathrm{s}$ using $M=10^4$ samples [Eq.~(\ref{eq: evidence})] and the estimated Bayesian evidence is in excellent agreement with that from nested sampling, implying any bias due to a lack of posterior probability mass coverage is negligible. Note that the quoted uncertainty in the evidence from importance sampling arises from the Monte Carlo variance of a single estimator; we repeated this estimate with 10 independent batches of $M=10^4$ samples each and found the spread in $\ln\mathcal{Z}'$ to be the same as the single-estimator uncertainty $\approx0.02$. Overall, variational inference with importance sampling effectively required $\approx10$ likelihood evaluations per posterior sample, compared to $10^2$--$10^3$ for nested sampling; without importance sampling, the overall efficiency of variational inference is uncapped as an arbitrary number of posterior samples can be drawn without evaluating the likelihood again after training.

\begin{table*}
\centering
\setlength{\tabcolsep}{10pt}
\renewcommand*{\arraystretch}{1.1}
\begin{tabular}{ccccc}
\hline 
\textbf{Algorithm} & \textbf{Likelihood evaluations} & \textbf{Approx. time} & \textbf{Evidence} $\ln\mathcal{Z}'$ & \textbf{Efficiency} \\
\hline
Nested sampling (default) & $1.1\times10^6$ & 25\,min & $-448.9\pm0.2$ & 0.4\% \\
Nested sampling (alternate) & $3.6\times10^5$ & 10\,min & $-449.0\pm0.2$ & 1\% \\
\hline
Variational inference & $10^4$ ($+10^4$) & 50\,s (+5\,s) & ($-448.9\pm0.02$) & (15\%) \\
Variational inference & $10^5$ ($+10^4$) & 3\,min (+5\,s) & ($-448.9\pm0.02$) & (3\%) \\
\hline
\end{tabular}
\caption{Summary statistics for inference runs on binary BHs through GWTC-3 using an NVIDIA A30 GPU. For variational inference, quantities in brackets are from the additional importance-sampling step.}
\label{tab: current stats}
\end{table*}

In Fig.~\ref{fig: current}, we show the posterior distributions inferred using nested sampling, compared to neural variational inference with and without smoothed importance sampling. To highlight any discrepancies where they are most likely to occur in the tails of the distributions, we show credible regions at the 99\% level, as well as 90\% and 50\%. Overall, the variational approximation matches nested sampling very well, capturing all of the correlations and nontrivial morphology in the posterior. There is good agreement even at the 99\% credible level, with only minor differences in some of the two-dimensional marginal distributions. In many instances, these are corrected by the smoothed importance weights, e.g., for the most discrepant marginal posterior in $\beta_\chi$. On the whole the variational posteriors with and without importance sampling are very similar, implying the initial trained normalizing flow is a sufficient representation of the posterior.

We also verify this by showing the corresponding inferred population-level distributions of the binary BH source parameters and their uncertainties. For primary BH mass $m_1$ and mass ratio $q$, we show the comoving merger rate densities $R(m_1)$ and $R(q)$ evaluated at redshift $z=0$. For spin magnitudes $\chi_i$ and tilts $\tau_i$, we show probability densities. We also show the evolution of the merger rate over redshift. As before, we include the central 50\%, 90\%, and 99\% credible regions. Again, nested sampling and variational inference agree well on the predicted BH distributions.

We also repeated variational inference but trained for $10^5$ steps (still with a batch size $B=1$). We similarly found $\varepsilon\approx0.3$, $\hat{k}\approx0.33$, and an identical $\ln\mathcal{Z}'$ (see Table~\ref{tab: current stats}), implying the shorter training run with $10^4$ steps is sufficiently converged. We compare posteriors from these runs in Appendix~\ref{app: Longer training runs}. Overall, these results indicate that neural variational inference is a promising tool for rapidly but accurately constraining the astrophysical populations of GW sources using real observations in current LVK data.

\subsection{Simulated catalog}
\label{sec: Simulated catalog}

We further test our inference approach with synthetic binary BH mergers, using the simulated catalog of Refs.~\cite{Vitale:2025lms, vitale_data_2025}. The same population models as above were used to generate sources, except that there is a correlation between the fraction $f(q)$ of BHs with aligned spins and their binary mass ratios $q$, given by $f(q) = f_\tau \big( e^{(q-0.1)^2} - 1 \big) / \big( e^{0.81} - 1 \big)$, where $f_\tau$ is now the fraction of sources with aligned spins at $q=1$. When performing inference, we neglect this correlation, which is a conservative choice in the sense that we can test whether the marginal spin distribution is correctly inferred even if we use an incorrect model. The true parameters of this population were chosen to be consistent with the constraints from current data (as in Fig.~\ref{fig: current}) and are given in Table~\ref{tab: future priors}, as are the priors we use for inference.

\begin{table}
\setlength{\tabcolsep}{3pt}
\renewcommand*{\arraystretch}{1.1}
\begin{tabular}{cccc}
\hline 
\textbf{Parameter} & \textbf{Description} & \textbf{Range} & \textbf{True} \\
\hline
$\alpha_m$ & $m_1$ spectral index & $[3,4]$ & 3.4 \\
$\beta_q$ & $q$ spectral index & $[0,2]$ & 1.1 \\
$f_m$ & $m_1$ Gaussian fraction & $[0,0.1]$ & 0.04 \\
$\mu_m$ & $m_1$ Gaussian location [$M_\odot$] & $30,35]$ & 34 \\
$\sigma_m$ & $m_1$ Gaussian width [$M_\odot$] & $[2,6]$ & 3.6 \\
$m_\mathrm{min}$ & Minimum BH mass [$M_\odot$] & $[4,6]$ & 5 \\
$m_\mathrm{max}$ & Maximum BH mass [$M_\odot$] & $[80,100]$ & 87 \\
$\delta_\mathrm{min}$ & Low-mass smoothing [$M_\odot$] & $[0,10]$ & 4.8 \\
$\delta_\mathrm{max}$ & High-mass smoothing [$M_\odot$] & $[0,10]$ & 0 \\
$\alpha_\chi$ & $\chi_i$ beta distribution shape $\alpha$ & $[1,3]$ & 1.67 \\
$\beta_\chi$ & $\chi_i$ beta distribution shape $\beta$ & $[1,6]$ & 4.43 \\
$f_\tau$ & Aligned-spin fraction & $[0,1]$ & $1^*$ \\
$\sigma_\tau$ & Aligned-spin $\cos\tau_i$ width & $[0.2,2]$ & $1.15^*$ \\
$\gamma$ & $z$ spectral index & $[1,4]$ & 2.73 \\
\hline
\end{tabular}
\caption{Parameters of the population model, the range of the uniform priors we use for inference, and the true underlying values for a simulated GW catalog containing 1599 binary BH mergers. $^*$Note: As the true population has a correlation between mass ratio $q$ and spin tilts $\tau_i$, these parameters do not directly correspond to those for the population model described in Sec.~\ref{sec: Likelihood and model}.}
\label{tab: future priors}
\end{table}

The catalog assumed a three-detector network of LIGO Hanford, LIGO Livingston, and Virgo \cite{LIGOScientific:2014pky, VIRGO:2014yos} with sensitivities as projected for the fourth observing run \cite{KAGRA:2013rdx, T2000012}. Signals were generated using the \textsc{IMRPhenomXP} waveform model \cite{Pratten:2020ceb}. Events were selected based on a detection threshold on the network match-filter signal-to-noise ratio (SNR) $>11$. The catalog contains full Bayesian parameter estimation with standard GW priors (see, e.g., \cite{KAGRA:2021vkt}) for 1599 detected events using the relative binning likelihood approximation in \textsc{Bilby} \cite{Ashton:2018jfp, Krishna:2023bug, Zackay:2018qdy, Cornish:2010kf}. We assume an observing time of ten years, such that the local merger rate is consistent with the current constraints in Fig.~\ref{fig: current} (note that this is not a realistic estimate for the number of mergers in the fourth LVK observing run or its duration, but was chosen to stress test the likelihood-based variational approach using a large catalog).

To estimate the population-level selection function in the likelihood of Eq.~(\ref{eq: likelihood}), we generated a custom set of $1.1\times10^7$ signals with $\mathrm{SNR}>11$. Combined with the $\approx1.6\times10^4$ parameter-estimation samples per event in the catalog, computing the likelihood requires evaluating the population model on $>3.6\times10^7$ samples. Analyzing catalogs with more events carries larger Monte Carlo variance $\mathcal{V}(\lambda)$ \cite{Essick:2022ojx, Talbot:2023pex}, hence the need for larger sample sizes; we relax the variance threshold to $V=4$ to aid sampling. The large sample count increases the computational cost of each likelihood evaluation, highlighting the need for sample-efficient inference algorithms. We perform a stringent test of neural variational inference by using a batch size $B=1$ and training for just $10^3$ steps, meaning the likelihood is evaluated only $10^3$ times during training. We again make $M=10^4$ draws from the trained posterior to use for importance sampling, which requires the same number of likelihood evaluations.

\begin{figure}
\centering
\includegraphics[width=0.9\columnwidth]{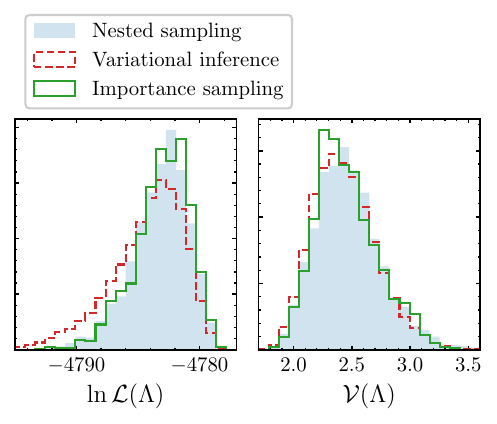}
\caption{The posterior distribution of log-likelihood values $\ln\mathcal{L}(\lambda)$ (left) and associated variances $\mathcal{V}(\lambda)$ (right) from a population analysis of a simulated catalog containing 1599 BH mergers inferred using nested sampling (shaded blue), variational inference (dashed red), and variational inference with smoothed importance sampling (solid green).}
\label{fig: extras-1599}
\end{figure}

As in Fig.~\ref{fig: extras}, we show the distributions for the likelihood evaluations given by $\ln\mathcal{L}(\lambda)$ on posterior draws and corresponding variances $\mathcal{V}(\lambda)$ for the simulated catalog in Fig.~\ref{fig: extras-1599}. Here, there is more discrepancy between the nested sampling and variational inference results due to the lower number of likelihood evaluations during training. Smoothed importance sampling suppresses the lower likelihood tail in the variational inference result, bringing the distribution more in line with nested sampling. In all cases, all posterior samples satisfy the imposed threshold $\mathcal{V}(\lambda) < V$. With only $10^3$ likelihood evaluations during training, the importance sampling efficiency is lower than before, around $\varepsilon\approx 10\%$, whereas the Pareto smoothing diagnostic $\hat{k}\approx0.53$ is larger, demonstrating that large $\hat{k}$ indeed correspond to worse variational fits.

We explore this further by presenting summary statistics in Table~\ref{tab: future stats}. We find similar timing and sample-efficiency ratios between nested sampling and variational inference as in Table~\ref{tab: current stats}, both having longer run times due to the increased computational cost of evaluating the likelihood function: a few hours for nested sampling and a few minutes for variational inference, which is promising for population studies on future catalogs with large numbers of events (though in practice such catalogs will contain sources with improved parameter estimates due to better detector sensitivities than considered here and waveform models with more detailed physical effects, such as higher modes \cite{Varma:2019csw, Pratten:2020ceb, Ramos-Buades:2023ehm} and eccentricity \cite{Islam:2021mha, Nagar:2024dzj, Gamboa:2024hli, Morras:2025nlp}). The evidence estimated via smoothed neural importance sampling is lower than that for nested sampling, suggesting that the variational posterior has not explored the full range of support; as mentioned in Sec.~\ref{sec: Importance sampling and evidence estimation}, this is a generic occurrence in variational inference. Despite this, the evidence estimates are broadly consistent and on the Jeffreys scale the differences are not statistically significant \cite{jeffreys1998theory}.

\begin{table*}
\setlength{\tabcolsep}{10pt}
\renewcommand*{\arraystretch}{1.1}
\begin{tabular}{ccccc}
\hline 
\textbf{Algorithm} & \textbf{Likelihood evaluations} & \textbf{Approx. time} & \textbf{Evidence} $\ln\mathcal{Z}'$ & \textbf{Efficiency} \\
\hline
Nested sampling (default) & $1.1\times10^6$ & 500\,min & $-4801.3\pm0.2$ & 0.4\% \\
Nested sampling (alternate) & $4.5\times10^5$ & 160\,min & $-4801.4\pm0.2$ & 1\% \\
\hline
Variational inference & $10^3$ ($+10^4$) & 4\,min (+4\,min) & ($-4801.9\pm0.03$) & (10\%) \\
Variational inference & $10^4$ ($+10^4$) & 25\,min (+4\,min) & ($-4801.6\pm0.01$) & (25\%) \\
\hline
\end{tabular}
\caption{Summary statistics as in Table~\ref{tab: current stats} but for inference runs on a simulated catalog of 1599 BH mergers.}
\label{tab: future stats}
\end{table*}

\begin{figure*}
\includegraphics[width=\textwidth]{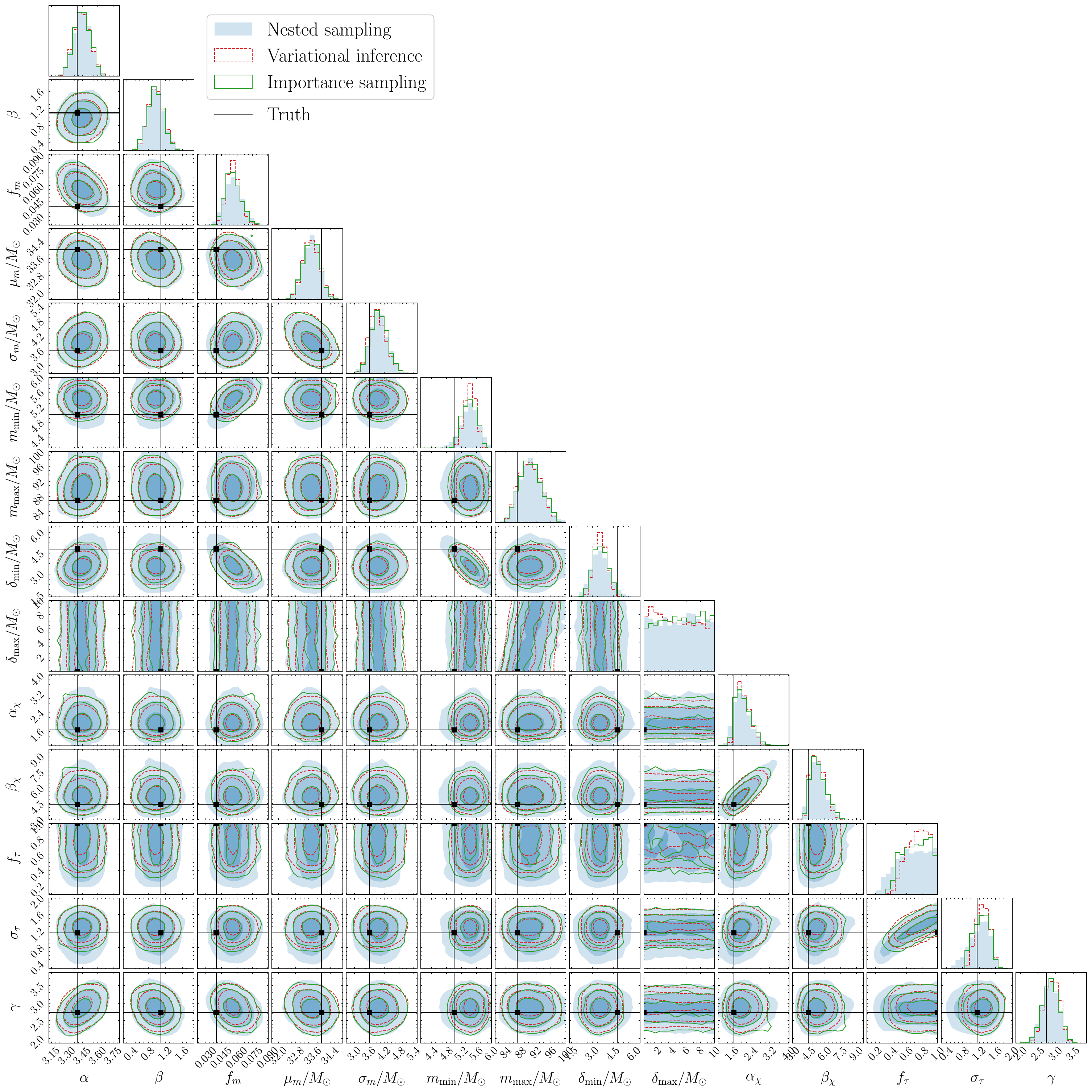}%
\llap{\raisebox{7.6cm}{%
    \includegraphics[width=0.6\textwidth]{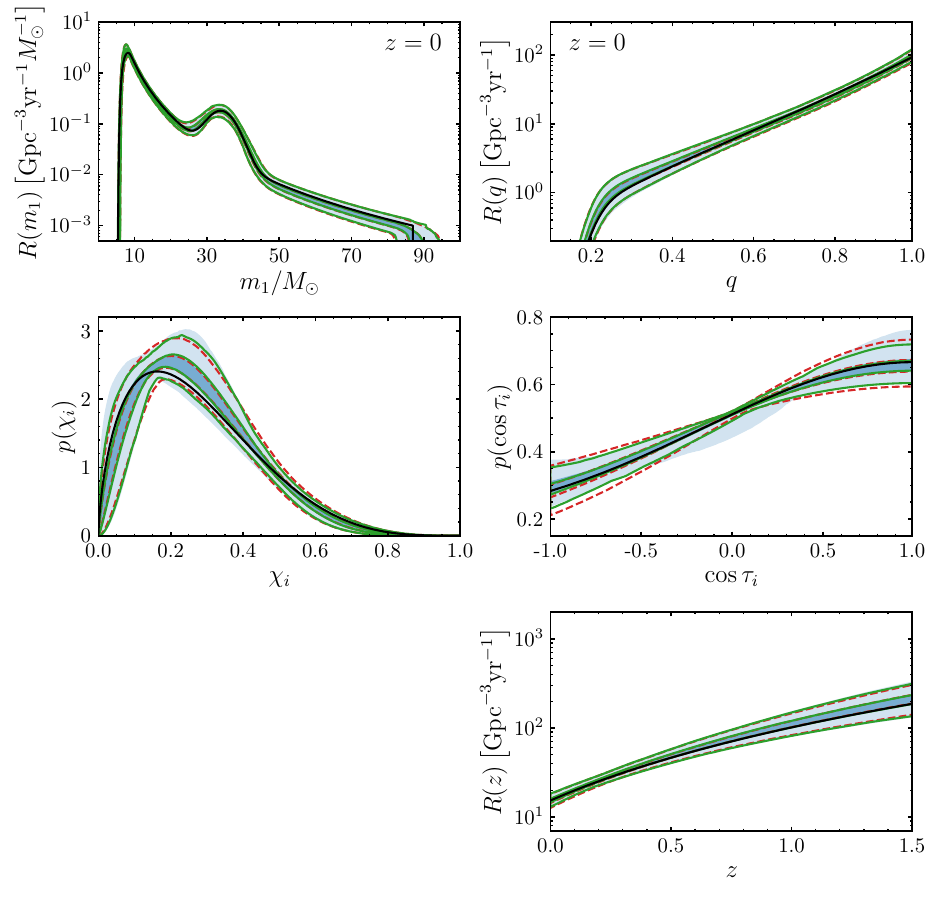}%
}}
\caption{As in Fig.~\ref{fig: current}, but for the simulated catalog of 1599 BH mergers. True parameters and source distributions are overplotted in black. For the posterior populations in the upper right, we only show the 50\% and 99\% credible regions for clarity.}
\label{fig: future}
\end{figure*}

The differences between posteriors from nested sampling and variational inference are visualized in Fig.~\ref{fig: future}. Several discrepancies are apparent at the 99\% credible level, especially in the population parameters that govern the BH spin distributions. However, much better agreement is observed in parameters governing the mass distribution. Combined with the likelihood variances, this implies that BH spin measurements drive uncertainty in estimating the population likelihood and thus impair robust inference. Importance sampling (with or without smoothing) is not able to correct for this misestimation as the initial variational posterior provides too few tail samples. This is apparent in the predicted distributions of spin magnitudes $\chi_i$ and tilts $\tau_i$.

Despite this, the overall conclusions about the inferred population are not impacted by these discrepancies and the true population is correctly recovered for all marginal population distributions within their posterior uncertainties; this includes the spin tilts, even though the model used for recovery neglects the underlying correlation with mass ratio in the injected population. This demonstrates that low Monte Carlo variance in the likelihood estimator (e.g., $\mathcal{V}(\lambda)<1$ \cite{Talbot:2023pex}) may not be entirely necessary for correct inference.

We also tested a longer training run with $10^4$ steps to confirm that this brings the variational posterior much more in line with the results of nested sampling; see Table~\ref{tab: future stats} and Appendix~\ref{app: Longer training runs}. In particular, unlike in Sec.~\ref{sec: Current catalog} and Table~\ref{tab: current stats}, we find that increasing the number of likelihood evaluations during training by a factor of 10 improves the overall efficiency (effective number of posterior samples per likelihood evaluation). However, we present results with so few likelihood evaluations during training to emphasize the usefulness of neural variational inference even in this restricted setting.

\section{Discussion}
\label{sec: Discussion}

Variational methods have previously been used for ground-based GW parameter estimation \cite{Gabbard:2019rde} and pulsar timing array data analysis \cite{Vallisneri:2024xfk}. Here, we showed the efficacy of neural variational inference---which trains a normalizing flow to approximate Bayesian posteriors by matching their shapes using likelihood evaluations---for population studies of GW catalogs. This requires only an input (differentiable) likelihood function, which is readily available in public code packages (e.g., \textsc{GWPopulation}~\cite{Talbot:2024yqw}), meaning the variational approach serves as an accessible inference algorithm and a drop-in replacement for common stochastic-sampling tools.

We found it requires at least an order of magnitude and up to three orders of magnitude fewer likelihood evaluations than established stochastic sampling algorithms such as nested sampling. This makes variational inference conducive to interactive development and comparison of population models, thereby accelerating astrophysical interpretation of GW catalogs. We demonstrated that accurate posteriors can be trained with as few as $\mathcal{O}(10^3$--$10^4)$ likelihood evaluations and a batch size of just $B=1$ to compute the variational loss function, implying that the training setup can be tailored to available hardware (e.g., GPU memory limitations) and to trade off accuracy against speed. Simply performing more training steps and thereby increasing the total number of likelihood evaluations was always sufficient to improve the variational approximation. Even if variational inference were to take as long as stochastic sampling runs, it still has the advantage of producing posterior distributions that, once trained, provide exact density evaluations and any number of posterior draws at negligible cost. A sizable contribution to the training time comes from JIT compilation, which we may be able to reduce with additional code optimization.

Trained variational posteriors could be useful for downstream applications, serving as rapid surrogates for likelihood functions when combining independent datasets (e.g., Ref.~\cite{Cousins:2025bas}). For example, rather than reanalyzing old GW catalogs every time new data is released, previous variational posteriors can be reused to encode the information already provided within the formalism of Bayesian updating. With the increased detection rate from detector upgrades and future GW observatories \cite{KAGRA:2013rdx, Baibhav:2019gxm, Broekgaarden:2023rta}, variational inference could be used for online updating of astrophysical population constraints as new events are detected. Independently of the variational approach, Pareto smoothed importance sampling and the $\hat{k}$ diagnostic could be useful tools to regularize the Monte Carlo approximations in the population likelihood and assess their convergence, without relying on the empirical variance \cite{chatterjee2018sample, JMLR:v25:19-556}.

Alongside recent advances in machine learning \cite{Gabbard:2019rde, Dax:2021tsq, Dax:2024mcn, Kolmus:2024scm} and gradient-based methods \cite{Wong:2023lgb, Wouters:2024oxj}, neural variational inference may be applicable to accelerate parameter estimation of individual GW sources, either by using the trained normalizing flows as posteriors directly or their transformations for more efficient (gradient-based) stochastic sampling \cite{hoffman2019neutra}. As a likelihood-based approach, variational inference can immediately take of advantage of accelerated likelihoods \cite{Cornish:2010kf, Smith:2016qas, Zackay:2018qdy, Morisaki:2021ngj, Morisaki:2023kuq} and waveforms \cite{Varma:2019csw, Khan:2020fso, Thomas:2022rmc, Edwards:2023sak}. However, there are potential issues with the training objective used, which can struggle to train posteriors with multimodalities or extended tails. In initial tests, we found that annealed posterior tempering was effective for target distributions with large separate modes (e.g., high-dimensional Gaussian mixtures), but struggled with deep potentials (e.g., the Rosenbrock function \cite{rosenbrock1960automatic, pagani2022n, Williams:2023ppp}). Other possible remedies include altering the training objective \cite{10.1145/3341156, Kolmus:2024scm} or its gradient estimator \cite{NIPS2017_e91068ff, vaitl2022gradients}.

Our work highlights the potential of machine learning to aid astrophysical inference in GW astronomy. In future work, we will continue to explore these applications. We hope the methods developed here and our implementation \cite{matthew_mould_2024_12770127} will aid analysts in designing population models for current and future studies of GW catalogs.

\section*{Acknowledgments}

We thank Lilah Mercadante, Ryan Abbott, Phiala Shanahan, Sofía Álvarez-López, Jack Heinzel, Cailin Plunkett, Jacob Golomb, Hui Tong, Aditya Vikaykumar, and the participants of the IFPU Focus Week workshop ``Emerging methods in GW population inference'' for helpful discussions.
M.M. is supported by the LIGO Laboratory through the National Science Foundation awards No. PHY-1764464 and No. PHY-2309200.
N.E.W. is supported by the National Science Foundation Graduate Research Fellowship Program under grant No. 2141064.
S.V. is partially supported by the NSF grant No. PHY-2045740.
This work made use of resources provided by subMIT at MIT Physics.
The authors are grateful for computational resources provided by the LIGO Laboratory and supported by National Science Foundation Grants No. PHY-0757058 and No. PHY-0823459.
This research has made use of data or software obtained from the Gravitational Wave Open Science Center (gwosc.org), a service of the LIGO Scientific Collaboration, the Virgo Collaboration, and KAGRA.
This material is based upon work supported by NSF's LIGO Laboratory which is a major facility fully funded by the National Science Foundation.
This work is supported by the National Science Foundation under Cooperative Agreement PHY-2019786 (The NSF AI Institute for Artificial Intelligence and Fundamental Interactions, http://iaifi.org/).
This work was partially supported by the MIT School of Science John W. Jarve (1978) Seed Fund.

\section*{Data Availability}

The data that support the findings of this article are openly available~\cite{KAGRA:2023pio, vitale_data_2025}.

\appendix

\section{LONGER TRAINING RUNS}
\label{app: Longer training runs}

Here, we present posterior distributions using a larger number of steps (and thus likelihood evaluations) to train the variational posteriors.

\begin{figure*}
\includegraphics[width=\textwidth]{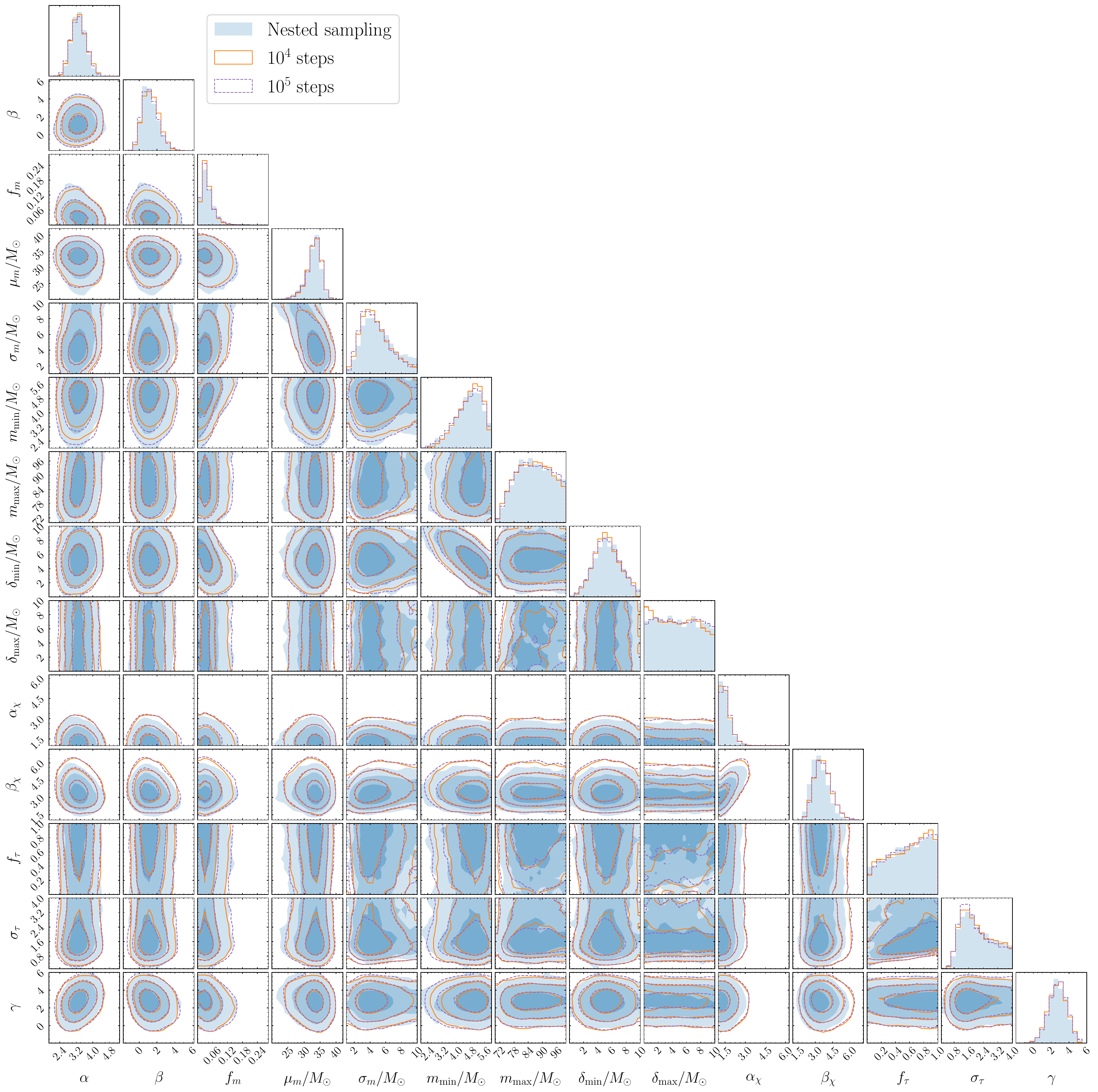}%
\llap{\raisebox{7.6cm}{%
    \includegraphics[width=0.6\textwidth]{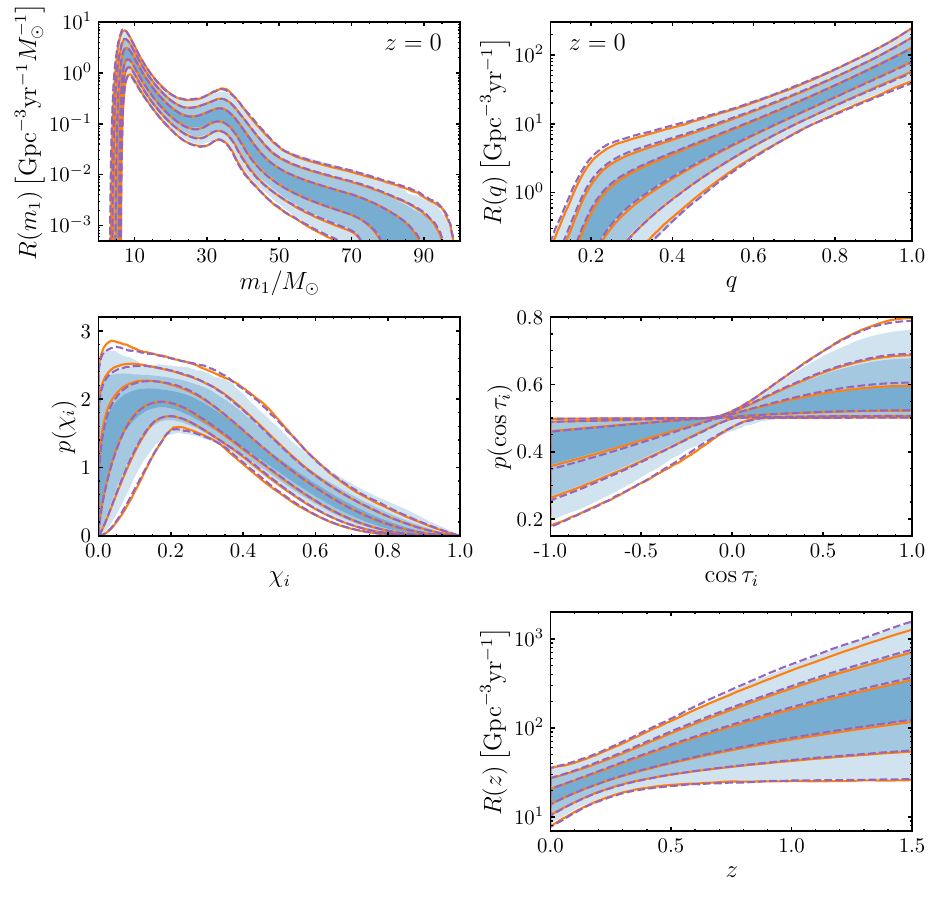}%
}}
\caption{Posterior distributions from population inference on GWTC-3, as in Fig.~\ref{fig: current}, but comparing nested sampling (shaded blue) to variational posteriors trained with $10^4$ (solid orange) and $10^5$ (dashed purple) likelihood evaluations, without importance sampling.}
\label{fig: steps current}
\end{figure*}

In Fig.~\ref{fig: steps current}, we compare the nested-sampling result for GWTC-3 to variational posteriors trained with $10^4$ and $10^5$ likelihood evaluations---but without (Pareto-smoothed) importance sampling, which tends to reduce differences between the normalizing flows alone. The two are essentially indistinguishable, again confirming that the shorter training run is sufficient to reach convergence. The one parameter that is visibly different is $\delta_\mathrm{max}$---the width of the high-mass tapering function---whose one-dimensional marginal posterior is basically the same as the prior and is therefore more easily influenced by sampling noise during training. Discrepancies that would be corrected by (Pareto-smoothed) importance reweighting can be seen when comparing to nested sampling. The summary statistics with importance sampling are given in Table~\ref{tab: current stats}.

\begin{figure*}
\includegraphics[width=\textwidth]{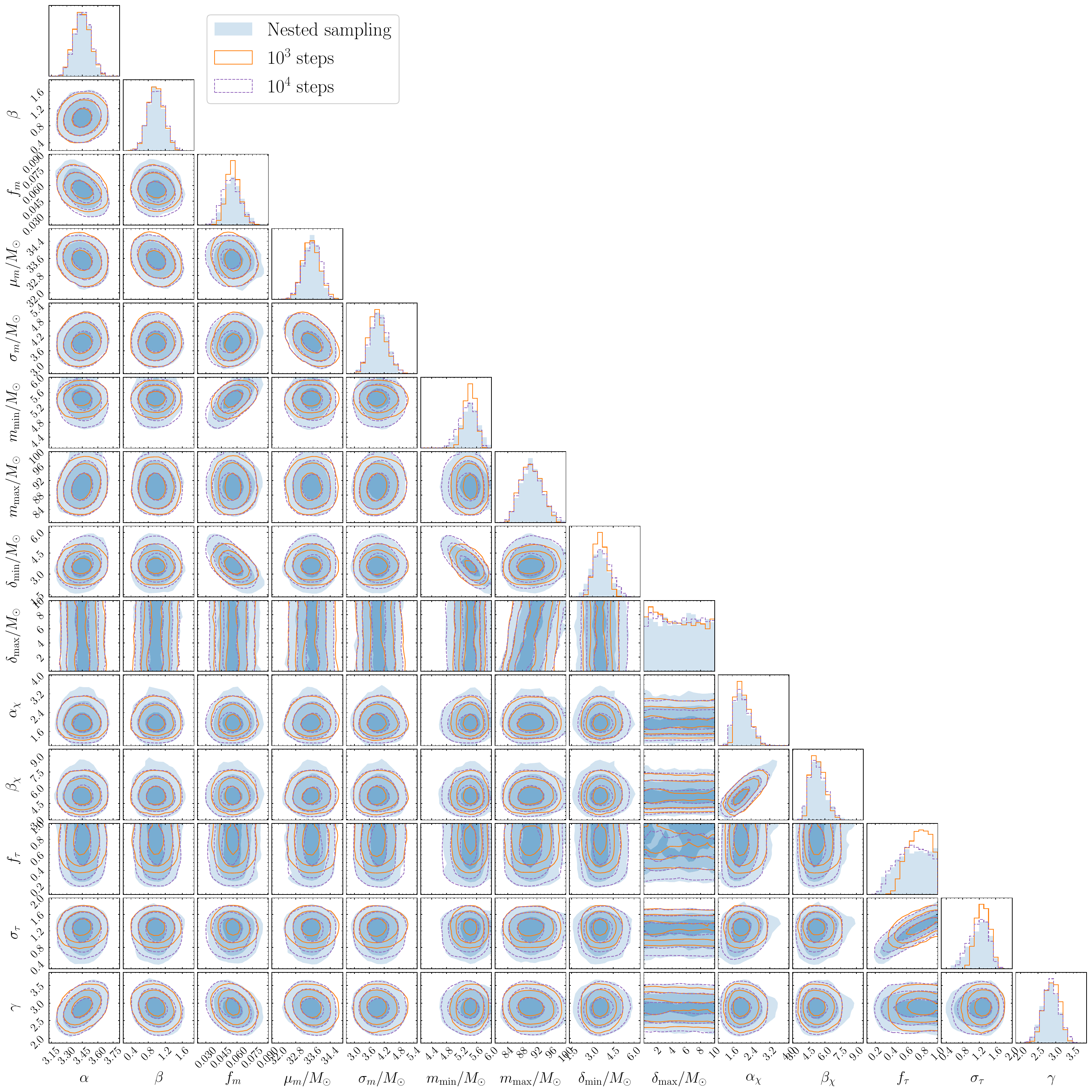}%
\llap{\raisebox{7.6cm}{%
    \includegraphics[width=0.6\textwidth]{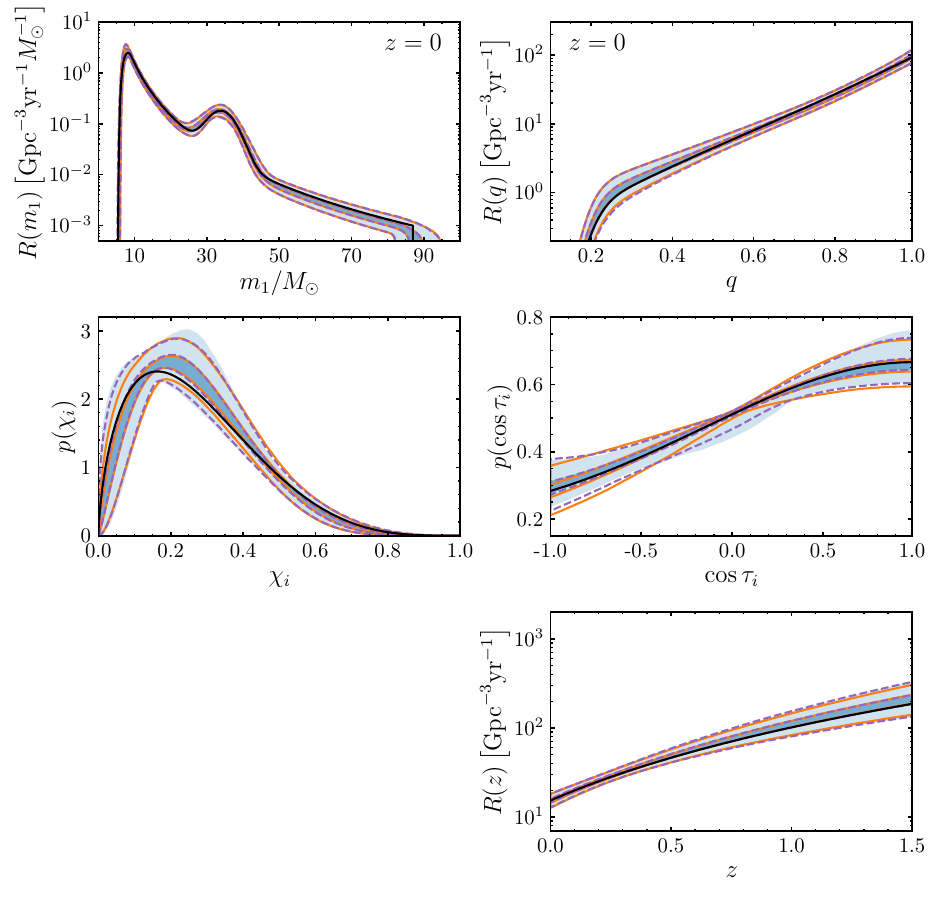}%
}}
\caption{Posterior distributions from population inference on the simulated catalog of 1599 events, as in Fig.~\ref{fig: future}, but comparing nested sampling (shaded blue) to variational posteriors trained with $10^3$ (solid orange) and $10^4$ (dashed purple) likelihood evaluations, without importance sampling.}
\label{fig: steps simulated}
\end{figure*}

In Fig.~\ref{fig: steps simulated}, we present the same plot but for the simulated catalog of 1599 events, with variational posteriors trained with $10^3$ and $10^4$ steps. In this case, the two variational posteriors differ much more, confirming that the shorter training run with $10^3$ likelihood evaluations---while producing reasonable posterior population distributions---is not fully converged. Increasing the number of training steps by a factor of 10 produces a variational posterior that matches nested sampling much better. The summary statistics with importance sampling are given in Table~\ref{tab: future stats}.

\bibliography{draft}

\end{document}